# Tailoring Hybrid Anomalous Hall Response in Engineered Magnetic Topological Insulator Heterostructures

*Peng Chen[1, ‡], Yong Zhang[1, ‡], Qi Yao[2,3, ‡,*], Fugu Tian[1], Lun Li[1], Zhengkun Qi[1], Xiaoyang Liu[1], Liyang Liao[4], Cheng Song[4], Jingyuan Wang[5], Jing Xia[5], Gang Li[2,3], David M. Burn[6], Gerrit van der Laan[6], Thorsten Hesjedal[7], Shilei Zhang[2, 3,*], and Xufeng Kou[1, 3,*]*

[1] School of Information Science and Technology, ShanghaiTech University, Shanghai 200031, China

[2] School of Physical Science and Technology, ShanghaiTech University, Shanghai 200031, China

[3] ShanghaiTech Laboratory for Topological Physics, ShanghaiTech University, Shanghai 200031, China

[4] Key Lab Advanced Materials (MOE), School of Materials Science and Engineering, Tsinghua University, Beijing 100084, China

[5] Department of Physics and Astronomy, University of California, Irvine, CA 92697, USA

[6] Magnetic Spectroscopy Group, Diamond Light Source, Didcot OX11 0DE, United Kingdom

[7] Clarendon Laboratory, Department of Physics, University of Oxford, Oxford OX1 3PU, United Kingdom

**ABSTRACT:** Engineering the anomalous Hall effect (AHE) in the emerging magnetic topological insulators (MTIs) has great potentials for quantum information processing and spintronics applications. In this letter, we synthesize the epitaxial $Bi_2Te_3$/MnTe magnetic heterostructures and observe pronounced AHE signals from both layers combined together. The evolution of the resulting hybrid AHE intensity with the top $Bi_2Te_3$ layer thickness manifests the presence of an intrinsic ferromagnetic phase induced by the topological surface states at the heterolayer-interface. More importantly, by doping the $Bi_2Te_3$ layer with Sb, we are able to manipulate the sign of the Berry phase-associated AHE component. Our results demonstrate the un-paralleled advantages of MTI heterostructures over magnetically doped TI counterparts, in which the tunability of the AHE response can be greatly enhanced. This in turn unveils a new avenue for MTI heterostructure-based multifunctional applications.

Magnetic topological insulators (MTIs), which integrate both topology and magnetism within one system, have greatly broadened the research scope of quantum materials [1]. In addition to the spin-momentum locked feature of the non-trivial topological surface states, the introduction of perpendicular magnetic moments can also polarize the electron spins, and hence may possess a better capability for the control of the spin states within the host systems [2-4]. Therefore, introducing MTIs into the field of spintronics defines a new trend of magnetic-based logic and memory applications, in essence, to read and write the binary information that is encoded by the electron spin using all-electrical means.

From a material point of view, a pronounced anomalous Hall effect (AHE) can be effectively generated in topological insulators (TIs) by magnetic doping[5-8], and its quantum version (i.e., the quantum anomalous Hall effect) has been realized where the dissipation-less chiral edge state is formed without applying the magnetic field [2-3, 8-11]. Alternatively, MTI can also be achieved by proximity coupling of a TI to a ferro-/antiferromagnetic material [12-20]. In such MTI heterostructures, the separation of topology and magnetism in different layers enables us to optimize each contribution independently. For instance, rich inter-/intra-layer exchange couplings could result in higher magnetic transition temperatures and desirable complex spin textures[18-20]. Meanwhile, because of the strong spin-orbit coupling nature, the giant spin-orbit torque (SOT) within the TI layer would permit highly efficient current-driven magnetization switching up to room temperature[21-23]. Accordingly, the structural engineering of MTI not only enriches the choice of materials that can be joint together, but also provides additional degrees of freedom to manipulate different physical orders.

Furthermore, in terms of device applications, while the read operation is evaluated by the difference between the '1' and '0' states from the electrical feedback (e.g., anomalous Hall resistance ratio) [24], the write performance is governed by the magnetization switching efficiency (e.g., SOT) that alters between the two spin states[25-27]. Given that both spin/anomalous Hall effects are closely related to the specific electronic structure in momentum space (i.e., the integration of the Berry phase curvature over the Brillouin zone)[28-29], we can, in principle, modulate the spin Hall angle (which in turns determines the strength of SOT) by precisely tuning the magnitude and polarity of the AHE response in the aforementioned MTI heterostructures.

In this letter, we report the growth of $Bi_2Te_3$/MnTe heterostructures by molecular beam epitaxy (MBE). The magnetic moments in the MnTe film can effectively couple with the interfacial $Bi_2Te_3$ topological surfaces states to form an additional ferromagnetic order, and the resulting hybrid AHE response can be quantified by a classical molecular field model. Concurrently, through the systematic variation of the $Bi_2Te_3$ layer thickness, we are able to separate the independent AHE components from each channel and identify that the induced magnetic phase is closely related to the Berry curvature at the interface. Besides, we demonstrate that the polarity of the TI-associated AHE component can be switched when the majority carrier within the TI layer changes from electrons to holes via counter-doping strategy. Such flexibility in controlling the AHE strength in our MTI heterostructures therefore opens up a multitude of opportunities to explore MTI-based spintronics devices.

Experimentally, the detailed sample structure was grown in a sequence of CrSe-MnTe-$Bi_2Te_3$ on the semi-insulating GaAs (111)B substrate using MBE, as shown in Figure 1a. As described in the previous reports[20], the insulating CrSe layer is incorporated as a buffer layer for the growth of the single-crystalline MnTe film, and we have deliberately designed a moderate post-annealing process to introduce long-range ferromagnetism in the MnTe layer (see Supporting Information S1). In order to make a solid comparison, we have fixed the thicknesses of the CrSe and MnTe layers for all the samples used in this study prior to the top $Bi_2Te_3$ layer growth. During the entire heterostructure growth process, surface-sensitive high-energy electron diffraction (RHEED) is used to monitor the *in-situ* growth dynamics. Figure 1b shows the RHEED patterns that are measured after the hetero-epitaxial deposition of each layer. The clear and sharp streaks confirm the two-dimensional growth mode for each species. Meanwhile, as labeled by the white dashed lines in Figure 1b, the *d*-spacing between the two first-order diffraction lines of the reciprocal RHEED patterns (i.e., reflect the in-plane lattice constant of the as-grown surface) always shrinks/expands to its new stable value immediately after one monolayer growth of each layer, thus signifying the sharp interface transition. Figure 1c displays the out-of-plane x-ray diffraction (XRD) data for the as-grown $Bi_2Te_3$/MnTe sample. It is seen that the major thin film peaks are dominated by the $Bi_2Te_3$ (*R3-m*) phase, and the rest of the peaks can be indexed by either the MnTe (*P63/mmc*) or the CrSe (*P63mc*) phase. Likewise, both the corresponding x-ray reflectivity (XRR) curve with multiple fringes and the cross-sectional transmission electron microscopy (TEM) image in Figure 1d manifest the single-crystalline property of the epitaxial MTI film with well-defined hetero-interfaces and negligible structural defects. Furthermore, we emphasize that the same film quality can be reproduced in all heterostructure samples with various TI layer thicknesses, and such stable material synthesis recipe guarantees the reliablity of our results discussed in this work.

The main objective of the presented study is to investigate the magneto-electrical properties and magnetic exchange coupling of the MTI hybrid system. In this respect, we first carried out the magneto-transport experiments on the eight quintuple-layer (QL) $Bi_2Te_3$/MnTe film and compared the field-dependent anomalous Hall ($R_{xy}$) result with the pure MnTe control sample at $T = 1.5$ K. As revealed in Figure 1e, both samples exhibit pronounced ferromagnetic order with perpendicular anisotropy. However, two additional symmetric $R_{xy}$ humps in the intermediate magnetic-field region show up in the hysteresis loop of the Bi$_2$Te$_3$/MnTe film. Noting that the square-like backgrounds and the relevant coercivity fields (~0.9 T) are almost identical for the two samples, we assume that while the consistent FM background stems from the MnTe layer, the $R_{xy}$ humps may imply the presence of intriguing interlayer magnetic interactions in our MTI heterostructure.

Inspired by the above discovery, we next conducted temperature-dependent measurements to quantitatively study the hybrid AHE in the 8 QL Bi$_2$Te$_3$/MnTe sample. It is seen from Figure 2a that with increasing temperature, the $R_{xy}$ humps shrink more markedly as compared to the background hysteresis loop. Specifically, the hump peak $\rho_{xy}^{total}$ to the saturated anomalous Hall resistance $\rho_{xy}^S$ (after subtracting the linear ordinary Hall background) ratio drops dramatically from 71% ($T$ = 3 K) to nearly zero for $T$ = 30 K, yet the square-like AHE response persists at a much higher temperature. Detailed magnetic phase diagrams for the AHE and magnetization ($M$) are shown in Figures 2b and 2c, respectively. Compared with the pure MnTe sample, it is concluded that both the saturated $\rho_{xy}^S$ and $M$ curves share the same temperature evolution trend (i.e., with the Curie temperature $T_C \approx 150$ K) of the MnTe background, while the pronounced AHE hump pocket at low temperature region (highlighted as the red area in Figure 2b) signifies its intrinsic relation with the top Bi$_2$Te$_3$ layer.

The observation of the symmetric anomalous Hall humps under moderate magnetic fields is reminiscent of the signature of the topological Hall effect (THE) where similar hysteresis line shapes have been discovered in several magnetic heterostructures with inversion symmetry breaking[30-32]. In those reported systems, the coupling between the itinerant electron spins and non-collinear local magnetic moments develops an emergent magnetic field that deflects the electrons. Nevertheless, we have carefully excluded such a THE-related origin in our case. First, since the non-collinear local magnetic moments in skyrmion-carrying systems are caused by the competitions among the Dzyaloshinskii-Moriya interaction, Heisenberg exchange interaction, and magnetic anisotropy, hence the THE feature generally occurs at a relatively high temperature regime close to $T_C$ [30-32]. However, as shown in Figures 2a-c, the occurrence of the $R_{xy}$ hump in our MTI heterostructures is found to strictly remain below 30 K and the hump intensity only discloses a monotonous temperature-dependent behavior, both of which are in contrast to the THE phase diagram. Furthermore, we have also performed systematic resonant elastic x-ray scattering (REXS) measurements at low temperatures, and the magnetic field-independent results show rather conventional ferromagnetic feature with no sign of the non-collinear spin configuration in real-space (see Supporting Information S2 for the discussion of the REXS data).

Alternatively, we notice that the complementary superconducting quantum interference device (SQUID) magnetization loop in Figure 2d shows a two-phase transition feature, and its characteristic transition field coincides with the AHE hump peak position. This finding may suggest that there are two magnetic phases undergoing magnetization reversal separately with opposite switching directions. Indeed, by applying the classical molecular field model of ferromagnetism (see Supporting Information S3) [33-34], we are able to separate the hybrid AHE of the 8 QL $Bi_2Te_3$/MnTe sample into two independent magnetic phases, as summarized in Figures 2e-f. Notably, while one extracted AHE loop ($\rho_{xy}^{MnTe}$) in Figure 2e perfectly reflects the original MnTe order, the other component ($\rho_{xy}^{TI}$) associated with the $Bi_2Te_3$ layer in Figure 2f shows a negative AHE polarity, namely the anomalous Hall resistivity saturates at a negative value when the system is magnetized along the +$c$-direction. Fundamentally, it is known that the intrinsic AHE is closely related to the Berry curvature of the occupied bands across the Fermi surface [28-29]. As proposed in the $SrRuO_3$ system with a band-crossing electronic structure, the AHE is a result of the integration of the Berry-phase curvature up to the Fermi level. Similarly, such a scenario may also apply for the surface states of $Bi_2Te_3$ at the heterolayer-interface, in which the sign of the AHE is governed by the relative position of the Fermi level (i.e., the band-filling point) with respect to the Dirac point.

To further address the observed negative-AHE physics and elucidate its intrinsic relation with the TI layer, we prepared a series of $Bi_2Te_3$/MnTe heterostructures by varying the $Bi_2Te_3$ thickness $d_{TI}$, for which Figure 3 lists the systematic results on three selected samples with $d_{TI}$ = 0 QL, 4 QL, and 16 QL. Comparing Figures. 3a-c, it can be clearly seen that the AHE hump intensity has a positive correlation with the top $Bi_2Te_3$ layer thickness, and a giant $\rho_{xy}^{total}/\rho_{xy}^{S}$ ratio exceeding 200% is realized in the 16 QL sample. In the meantime, the consistency of the MnTe-related FM background is repeatedly verified through the over-lapped $H_c^{AHE} - T$ traces (Figure 3d) and the magnetic phase diagrams (i.e., in order to make a fair comparison, we chose the same color code in Figures 3a2-c2) of all samples; on the contrary, the strength and transition temperature of the other magnetic phase at low temperatures are found to be strongly sensitive to the TI layer in the engineered heterostructures. Such thickness-dependent behaviors again highlight the indispensable role of the TI surface states in determining the AHE response within the hybrid system. Particularly, when the top TI layer is thin (e.g. the 4 QL sample), the injection of the magnetism and the associated magnetic coupling at the interface are strongly influenced by the hybridization between the top and bottom surfaces; yet once a larger portion of current conducts along the fully-developed topological surface states in the thicker TI layer at low temperatures (e.g., $T$ < 30 K), the resulting enhanced spin accumulation could contribute to the more pronounced humps (i.e., the induced AHE component $\rho_{xy}^{TI}$), as shown in Figures 2a and 3c1. Besides, we need to point out the temperature-dependent Hall resistance curves in both the 4QL and 16 QL $Bi_2Te_3$/MnTe samples exhibit a hook-like shape where $\rho_{xy}^{Total}$ reverses the trend and starts to increase when $T < 10$ K, as marked by the black arrows in Figure 3e. Compared with the MnTe control sample, it may indicate a rather mixed transport feature caused by the two un-correlated magnetic phases in our system.

Finally, we show the use of material engineering for the manipulation of the AHE in our MTI heterostructures. Given that the Berry curvature is highly sensitive to the Fermi level position and the details of the electronic structure, we further adopted the Sb counter-doping method [35-36] to prepare another 16 QL $p$-type $(Bi_{0.25}Sb_{0.75})_2Te_3$/MnTe film (see the raw transport data in Supporting Information S4) [37-38]. Strikingly, when the Fermi level is tuned below the Dirac point, the AHE humps in the $n$-type $Bi_2Te_3$/MnTe heterostructures give way to the appearance of a pair of symmetric dips during the magnetization reversal on the overall $\rho_{xy}^{Total}$ contour at 1.5 K, as illustrated in Figures 4a and 4b. Following the same fitting procedure, we manage to decompose the two-phase transition loop of the $(Bi_{0.25}Sb_{0.75})_2Te_3$/MnTe sample, and the results are presented in Figures 4c and 4e. It is noted that the MnTe channel always provide the identical AHE hysteresis line-shape with the same coercivity field among the two samples (Figure 4d); on the other hand, the AHE component contributed from the $p$-type TI channel displays a positive polarity, that is the saturated $R_{xy}$ reaches to the positive state under the $B > 0$ condition.

Interestingly, there is no report so far that the AHE sign can be inverted by varying the Fermi level position in magnetically-doped TIs. For example, the Mn/Cr-doped $(Bi_xSb_{1-x})_2Te_3$ systems always exhibit the negative/positive AHE sign regardless of the majority carrier type[7, 39-40]. Consequently, we believe that the ferromagnetic origin of our MTI heterostructures is distinct to the carrier-mediated Ruderman–Kittel–Kasuya–Yosida (RKKY) interaction which is dominant in uniform MTI systems. Instead, due to the giant spin susceptibility of the band-inverted topological surface states[2], the un-compensated $Mn^{3+}$ $d$-orbital electrons could effectively couple with the local band electrons of the TI surface near the interface to form a new magnetic order, and its strength thereafter would depend on the specific band structure (i.e., Berry curvature) at the Fermi level. Moreover, the advantage to tailor both amplitude and sign of the AHE response in our MTI heterostructures may facilitate the SOT-related device applications. Owing to the same origin of the spin/anomalous Hall effects[29], the direction of the SOT within the TI layer is hence determined by the AHE polarity. Under such circumstance, one may envisage that with further electric field-controlled capability, the top gate can be used to finely tune the Fermi level position across the Dirac point of the top $(Bi_xSb_{1-x})_2Te_3$ layer in the MTI heterostructure. Therefore, the relevant SOT-driven magnetization switching could be effectively manipulated in an all-electrical manner.

In conclusion, the $Bi_2Te_3$/MnTe heterostructures offer a great flexibility for the separation of various phases, as well as the unique ability to modulate the AHE signal via structural engineering, which is significant for further spin/magnetic manipulation in magnetic heterostructures. This work can lead to new strategies for MTI-related spintronics applications, such as tunable AHE sensor and SOT-based memory devices. Also, our results stimulate further investigations into quantifying the correlation between the intrinsic Berry curvature and the topological surface states-associated AHE. These will rely on systematic energy band density functional theory (DFT) studies and additional field-effect transistor device fabrication, which could lead to novel electrically controlled functionalities.

**EXPERIMENTAL DETAILS**

*Sample Fabrication*: The $Bi_2Te_3$/MnTe films were grown on epi-ready semi-insulating GaAs(111)B substrate by co-evaporation in MBE (base pressure $1 \times 10^{-10}$ mbar). Prior to sample growth, the GaAs substrate was pre-annealed at 570°C under the Se-protected environment in order to remove the native oxide surface. The growth temperatures for the CrSe and MnTe layers were 180°C and 370°C, respectively. Then the $Bi_2Te_3$ and $(Bi_{0.25}Sb_{0.75})_2Te_3$ films were grown at 200°C to minimize the Mn diffusion into the TI layers. During MBE growth, high-purity Cr, Mn, and Bi atoms were evaporated from standard Knudsen cells, while Se, Sb, and Te were evaporated by standard thermal cracker cells. Meanwhile, the beam flux monitor was used to measure element flux rate and RHEED was applied to monitor the real-time growth conditions. After sample growth, TEM was used to verify the sample quality. Furthermore, x-ray diffraction and reflectivity were preformed to measure the crystal structure and to calibrate the sample thickness.

*Transport Measurement*: The thin film heterostructures were manually etched into a six-probe Hall bar geometry with typical dimensions $2 \times 1mm^2$. The electrodes were made by welding indium shots onto the contact areas of the thin film. The magneto-transport measurements were performed with a He-4 refrigerator (Oxford TeslatronPT system). Several experimental variables such as temperature, magnetic field, and lock-in frequency were varied during the measurements. Multiple lock-in amplifiers and Keithley source meters (with an AC excitation current of $I = 1$ µA) were connected to the samples to enable the precise four-point lock-in experiments.

**FIGURES**

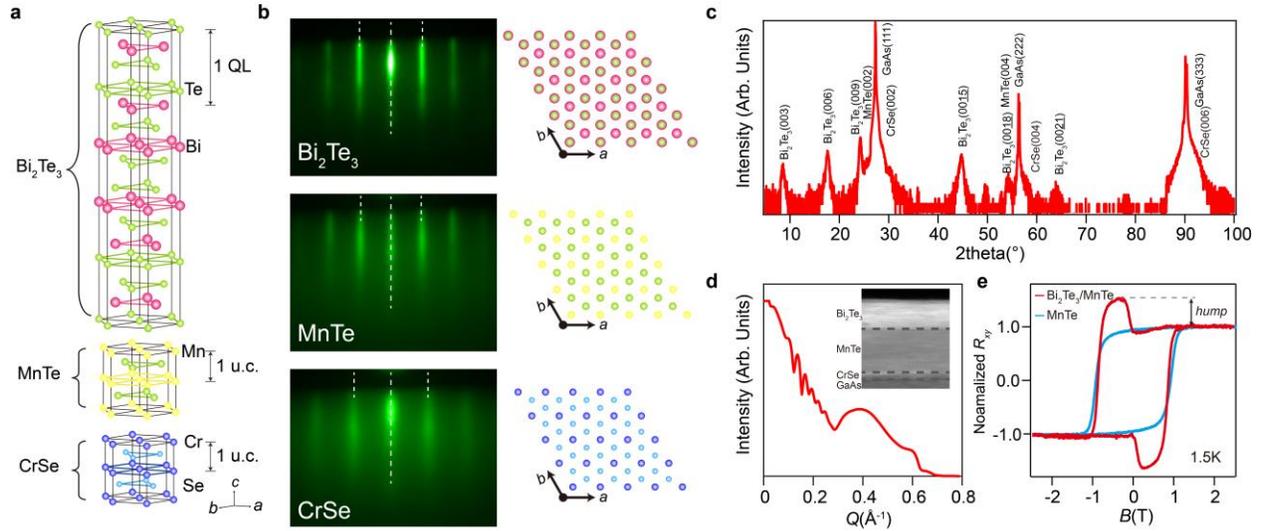

**Figure 1.** Structure characterizations of the MBE-grown $Bi_2Te_3$/MnTe heterostructures. (a) Illustration of the crystal structure of the $Bi_2Te_3$/MnTe/CrSe layer stack. CrSe is chosen as the buffer layer for the subsequent growth of the MnTe layer. (b) *in-situ* RHEED patterns of CrSe, MnTe, and $Bi_2Te_3$ during sample growth and the corresponding two-dimensional atomic configurations. (c) X-ray diffraction spectrum and (d) x-ray reflectivity pattern of the heterostructure sample. The inset shows the relevant cross-sectional TEM image, which unveils the single-crystalline structure with atomically flat heterolayer-interfaces. (e) Normalized field-dependent anomalous Hall resistance loops of the 8 QL $Bi_2Te_3$/MnTe and the pure MnTe films assigned by the red and cyan lines, respectively. At $T$ = 1.5 K, two symmetric humps appear between the two fully-magnetized states in the $Bi_2Te_3$/MnTe sample.

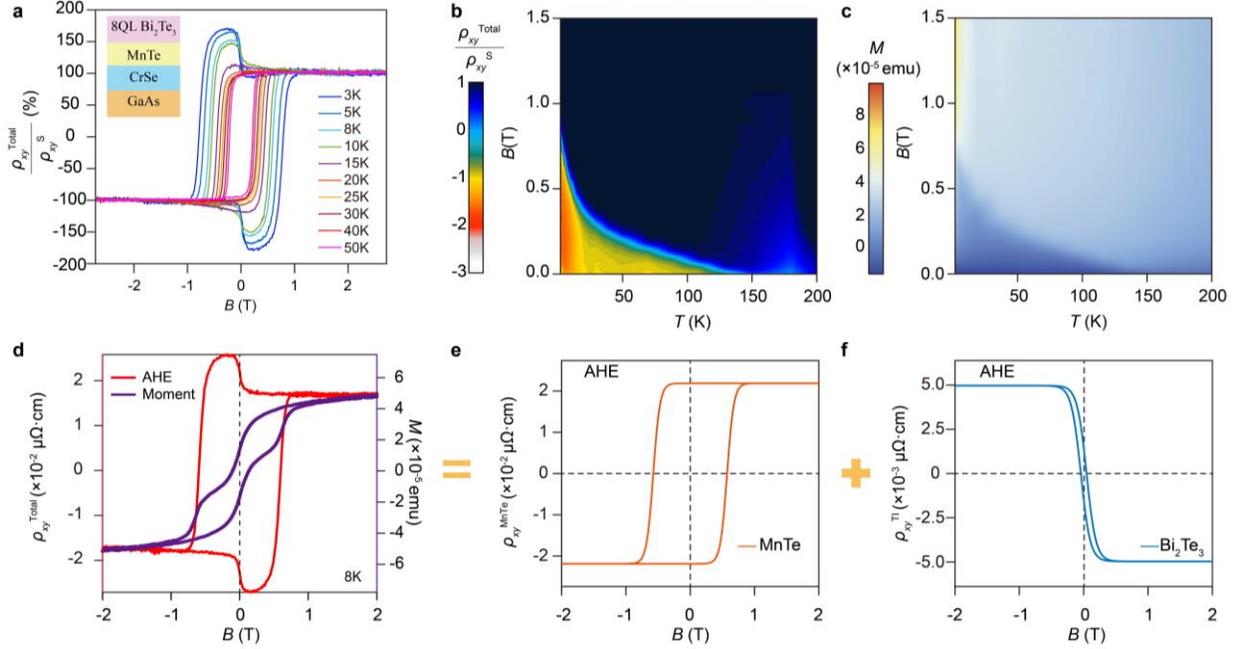

**Figure 2.** Temperature-dependent magneto-transport results and magnetic phase diagram of the 8 QL Bi$_2$Te$_3$/MnTe heterostructures. (a) Field-dependent hybrid AHE loops at different temperatures ranging from 3 K to 50 K, with $\rho_{xy}^{total}$ being the total Hall resistance and $\rho_{xy}^{S}$ the saturation value at high magnetic fields after subtracting the linear background. (b)-(c). Temperature vs. field phase diagrams for AHE and magnetization, respectively. (d) Comparison of the SQUID (purple line) and anomalous Hall (red line) hysteresis loops at 8 K. The hump feature of $\rho_{xy}^{total}$ coincides with the dips of the SQUID data in the same magnetic field region. The overall hybrid AHE curve in (d) can be separated by two contributions from the (e) MnTe and (f) Bi$_2$Te$_3$ layers.

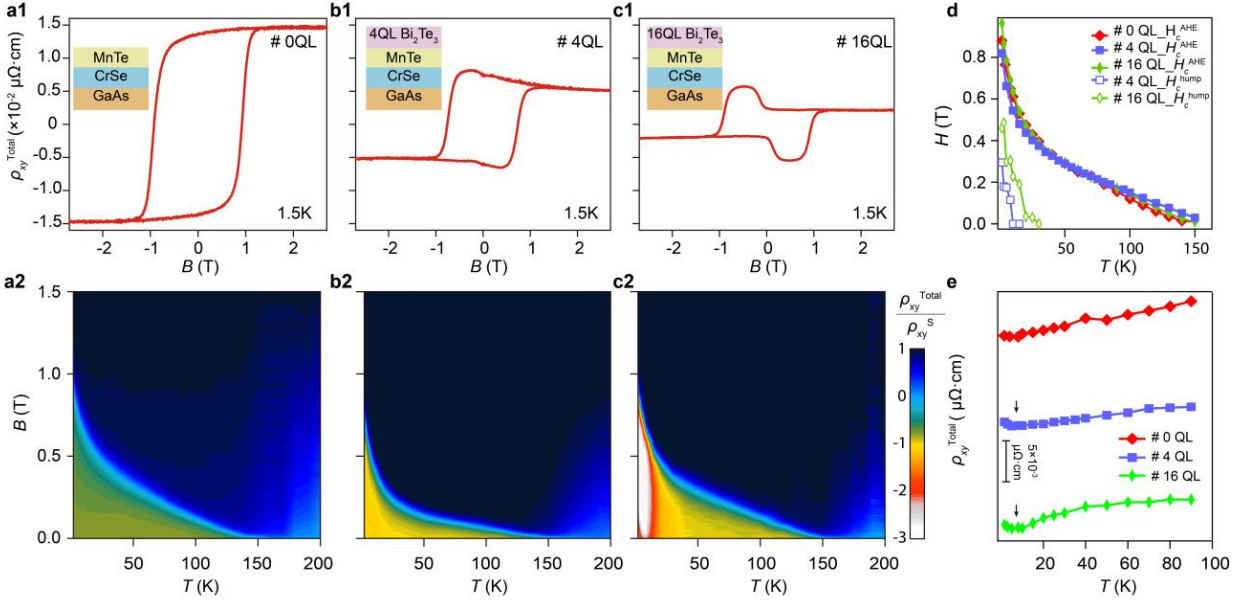

**Figure 3.** Evolution of the anomalous Hall humps in the engineered $Bi_2Te_3$/MnTe heterostructures with different TI layer thickness of 0 QL, 4 QL, and 16 QL. (a1-c1) Magnetic field-dependent hybrid anomalous Hall results at $T$ = 1.5 K. The hump magnitude becomes more pronounced with increasing $Bi_2Te_3$ layer thickness. (a2-c2) The corresponding magnetic phase diagrams of the three samples using the same color code. The giant enhancement of the $R_{xy}$ hump intensity can be observed clearly at low temperatures as the $Bi_2Te_3$ film thickness increases. (d) The summary of the temperature-dependent $H_c^{AHE}$ and $H_c^{hump}$ for MTI heterostructures with various $Bi_2Te_3$ thicknesses. (e) The comparison of temperature-dependent $\rho_{xy}^{total}$ for the three samples. The hook-like shape in both MTI samples indicated by black arrows around $T$ = 10 K may imply the presence of multiple magnetic phases in the system.

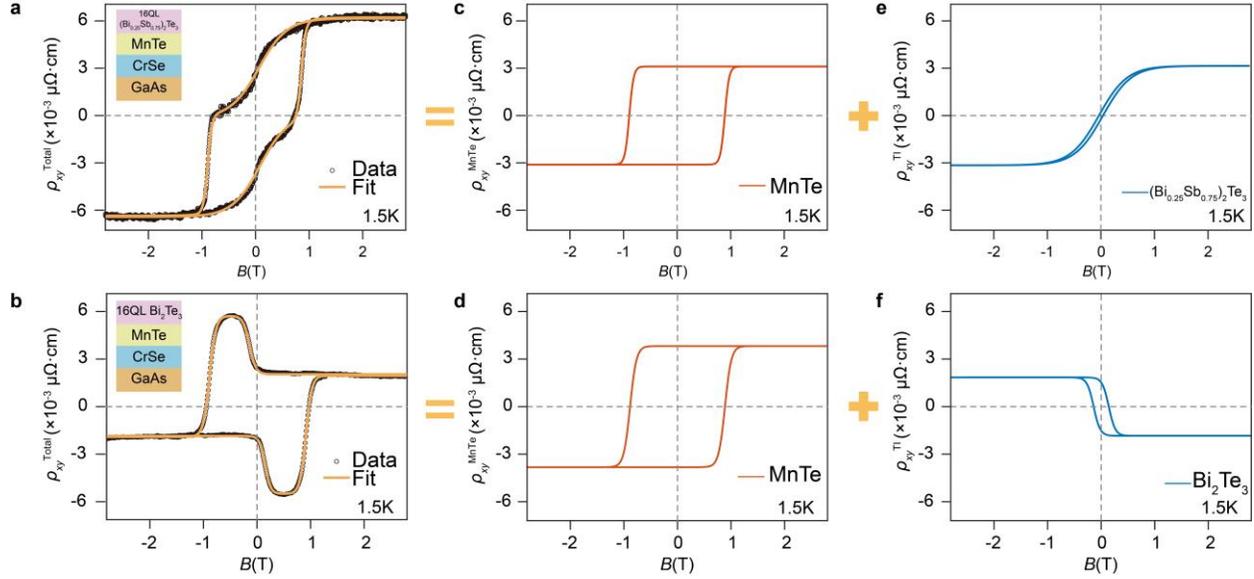

**Figure 4.** The manipulation of the AHE response in $(Bi_xSb_{1-x})_2Te_3$/MnTe systems. (a) – (b) Hybrid AHE results of the (a) *p*-type 16 QL $(Bi_{0.25}Sb_{0.75})_2Te_3$/MnTe and (b) *n*-type 16 QL $Bi_2Te_3$/MnTe [recaptured from Figure 3]. The black, open dots represent the measured data, while the fitting results are indicated by the khaki solid lines. The hybrid AHE curves in both samples can be decomposed into the MnTe [the orange lines in (c)-(d)] and the $Bi_2Te_3$ [the blue lines in (e)-(f)] contributions.

## AUTHOR INFORMATION

**Corresponding Author**

*Email: (Q.Y.) yaoqi@shanghaitech.edu.cn.

*Email: (S.L.Z.) zhangshl1@shanghaitech.edu.cn.

*Email: (X.F.K.) kouxf@shanghaitech.edu.cn.

**Author Contributions**

‡ P.C., Y.Z., Q.Y. contributed equally to this work.

## ACKNOWLEDGMENTS

We gratefully acknowledge experimental support from Drs. J. Shen, X. Wang, and helpful discussions with Drs. Q. M. Shao, L. Pan, and G. Yin. We acknowledge Drs. Q. Yang and W. Y. Liu in 'CℏEM', SPST of ShanghaiTech University (#EM02161943). This work is sponsored by the National Key R&D Program of China under the contract number 2017YFB0405704, National Natural Science Foundation of China (Grant No. 61874172), and the Major Project of Shanghai Municipal Science and Technology (Grant No. 2018SHZDZX02). We also acknowledge Diamond Light Source for time on beamline I10 under proposal MM21875 and MM23895. Q.Y. acknowledges the support from the Shanghai Sailing Program (Grant No. 19YF1433200) and National Natural Science Foundation of China (Grant No. 11904230). S.L.Z. and T.H. acknowledge funding from Engineering and Physical Sciences Research Council (EP/N032128/1). X.F.K. acknowledges the support from the 1000-Young talent program of China and the Shanghai Sailing program (Grant No. 17YF1429200).